\documentstyle[12pt]{article}

\begin{document}

\sloppy

\begin{flushright}{UT-701\\ Nov '95}\end{flushright}

\vskip 1.5 truecm

\centerline{\large{\bf On gaugino condensation }}
\centerline{\large{\bf in the effective theory}}
\vskip .75 truecm
\centerline{\bf Tomohiro Matsuda}
\vskip .4 truecm
\centerline {\it Department of Physics, University of Tokyo}
\centerline {\it Bunkyo-ku, Tokyo 113,Japan}
\vskip 1. truecm

\makeatletter
\@addtoreset{equation}{section}
\def\theequation{\thesection.\arabic{equation}}
\makeatother

\vskip 1. truecm

\begin{abstract}
\hspace*{\parindent}
We analyze the gaugino condensation in the effective
theory for N=1 SU(N) Supersymmetric QCD with $N_{f}$ flavors.
It is known that taking the vacuum expectation value of the 
matter field to be infinite,
we can show that gaugino condensation can occur.
At such a limit we should consider only pure  supersymmetric
Yang-Mills theory.
But when we include an interaction term of order
$O(\frac{1}{v})$, the situation can change.
We analyze the effect of this interaction term and examine the gaugino
condensation in the low energy Yang-Mills theory by using the scheme
of Nambu-Jona-Lasinio.

\end{abstract}

\section{Introduction}
\hspace*{\parindent}
\hspace*{\parindent}
If supersymmetry plays a role in low energy physics,
it is presumably broken.
When we think about dynamical breaking of supersymmetry,
we should necessarily consider the dynamical effect,
like gaugino condensation.
In ordinary gauge theories many attempts have been made 
to analyze the effect of 
fermion condensation, Nambu-Jona-Lasinio method is one of them.
Naively, NJL method seems difficult to  apply to supersymmetric 
models because there is a well-known non-renormalization theorem,
or because of the difficulty of introducing four-fermion interaction
and auxiliary field in a supersymmetric way.
 
In this paper we consider an intermediate scale effective Lagrangian of
N colors and $N_{f}$ flavors SQCD \cite{dine}.
This theory have a flat direction.
If matter field develops expectation value $v$ on this direction,
SU(N) gauge symmetry is broken,
and the low energy Lagrangian has two sectors(the kinetic term
for the pure Yang-Mills and Nambu-Goldstone), 
and their interaction term of order
O(1/$v$).
Taking the vacuum expectation value of the matter field
to be infinite, it becomes that what we should concern is 
the supersymmetric Yang-Mills theory without matter.
This theory is simple.
We can use instanton calculation at asymptotic(weak) region or
can use composite field effective Lagrangian in the strong
region, both of them coincide
to suggest the gaugino condensation.
However, when we include the interaction term which is responsible for
the anomaly of the original theory, the situation changes.
The instanton calculation is not responsible for the 
dynamical generation of F-term\cite{dine}, and the composite field analysis
becomes less trivial\cite{nills}.

So reconsidering this effective Lagrangian
from another point of view seems important.
We analyze the effect of the interaction term using the scheme of
Nambu-Jona-Lasinio.
Disappointingly, gaugino condensation is reliably observed
only when we take $N_{f}$ and $N$ to be enough large, but it is still
important to show that the interaction term can induce 
non-trivial effect on gaugino condensation. 

To derive this mass-gap equation, we used a tadpole method\cite{miller}.
The tadpole method is useful in discussing the quantum effect for 
supersymmetry breaking parameters because we can explicitly separate the vacuum
expectation values of fields and that we can deal with supersymmetry 
breaking parameter $F$ explicitly.
We should also note that we are not dealing with the original Lagrangian,
but with intermediate effective theory which seems to be important
from a phenomenological point of view.

\section{Supersymmetric QCD and Nambu-Jona-Lasinio method}
\hspace*{\parindent}  
In this section we mainly follow the argument of ref.\cite{dine}.

Our starting point is a Lagrangian with a gauge group $SU(N)$ and with 
$N_{f}$ flavors of quarks.
The $N_{f}$ quark flavor corresponds to $N_{f}$ chiral fields of the $N$
representation, $Q^{ir}(i=1,...,N;r=1,...,N_{f})$ and $\overline{N}$
representation, $\overline{Q}_{ir}$.
These superfields can be written with component fields as
\begin{equation}
  \label{component}
  \left\{
    \begin{array}{l}
      Q^{ir}=\phi^{ir}+\theta^{\alpha}\psi_{\alpha}^{ir}+\theta^{2}F^{ir}\\
      \overline{Q}_{ir}=\overline{\phi}_{ir}+\overline{\theta}_{\dot{\alpha}}
      \psi^{\dot{\alpha}}_{ir}+\overline{\theta}^{2}F_{ir}
    \end{array}
  \right.
\end{equation}
The gauge fields $A^{a}_{\mu}(a=1,...,N^{2}-1)$ are included in vector 
multiplets $V^{a}$ accompanied by their super-partners, gauginos $\lambda^{a}$
and auxiliary fields $D^{a}$.
The total theory is given by
\begin{eqnarray}
  \label{lag}
  L&=&\frac{1}{g^{2}}\left[\int d^{4}\theta Q^{+}e^{V}Q+\overline{Q}e^{V}
\overline{Q}^{+}+\int d^{2}\theta W^{\alpha a}W_{\alpha}
^{a}+h.c.\right]
\end{eqnarray}
Classically, this theory has a global $U(N_{f})_{Left}\times 
U(N_{f})_{Right}\times U(1)_{R}$ symmetry.
The $U(N_{f})_{Left}\times U(N_{f})_{Right}$ symmetry is just
like that of ordinary QCD, corresponding to separate rotation of the 
$Q$ and $\overline{Q}$ fields.
The symmetry $U(1)_{R}$ is a R-invariance, a symmetry under which 
the components of a given superfield transform differently.
This corresponds to a rotation of the phases of the grassmannian 
variables $\theta^{\alpha}$,
\begin{equation}
  \label{r}
  \left\{
  \begin{array}{lll}
    \lambda&\rightarrow&e^{i\alpha}\lambda\\
    \psi&\rightarrow&e^{i\alpha}\psi\\
    \overline{\psi}&\rightarrow&e^{i\alpha}\overline{\psi}
  \end{array}
  \right.
\end{equation}
or
\begin{equation}
  \left\{
    \begin{array}{lll}
      W_{\alpha}(\theta)&\rightarrow&e^{-i\alpha}W_{\alpha}(\theta 
               e^{i\alpha})\\
      Q(\theta)&\rightarrow&Q(\theta  e^{i\alpha})\\
      \overline{Q}(\theta)&\rightarrow&\overline{Q}(\theta  e^{i\alpha})
    \end{array}
  \right.
\end{equation}
Just as in ordinary QCD, some of these symmetries are explicitly 
broken by anomalies.
A simple computation shows that the following symmetry,
which is a combination of ordinary chiral $U(1)$ and $U(1)_{R}$
symmetry, is anomaly-free.
\begin{equation}
  \left\{
    \begin{array}{lll}
      W_{\alpha}(\theta)&\rightarrow&e^{-i\alpha}W_{\alpha}(\theta 
               e^{i\alpha})\\
      Q(\theta)&\rightarrow&e^{i\alpha(N-N_{f})/N_{f}}
               Q(\theta  e^{i\alpha})\\
      \overline{Q}(\theta)&\rightarrow&e^{i\alpha(N-N_{f})/N_{f}}
               \overline{Q}(\theta  e^{i\alpha})
    \end{array}
  \right.
\end{equation}
From now on, we call this non-anomalous global symmetry  $U(1)_{R'}$.

If $N_{f}<N-1$, the gauge group is not completely broken.
Moreover, we can see that instanton cannot generate a superpotential
in this case, so considering another non-perturbative effect 
in this model seems important.

For simplicity, we consider the case: $SU(N)$ gauge group is
broken to the pure $SU(N-N_{f})$.
The low-energy theory consists of two parts: Kinetic term
for the unbroken $SU(N-N_{f})$ gauge interactions and kinetic terms for the 
various massless chiral fields.

In addition to these terms, we should include higher dimensional
operators.
A dimension-five operator, in general, is generated at the
one-loop level.
This can be obtained by the renormalization of the effective
coupling:
\begin{equation}
  \label{geff}
  L=\frac{1}{g^{2}}\left[1+\frac{g^{2}}{32\pi^{2}}N_{f}ln
  \left(\frac{\phi}{M}\right)\right]W^{\alpha}W_{\alpha}
\end{equation}
$\phi$ must be chosen to be invariant under all non-R
symmetries.
The non-anomalous R-symmetry of the original theory must be realized
in the effective low-energy Lagrangian by the shift
induced by $\phi$.
That determines the R-charge of $\phi$ to be $(N-N_{f})/N_{f}$.

Let us analyze the effect of the dimension-five operator.
This term includes the interaction of the form:
\begin{equation}
  \label{int}
  \frac{g^{2}N_{f}}{32\pi^{2}}\frac{F_{\phi}}{<\phi>}\lambda\lambda
\end{equation}
Using the tadpole method\cite{miller},we can have:
\begin{eqnarray}
  \label{tadp}
  \frac{\partial V}{\partial F}&=&\chi
  \lambda\lambda+F^{*}\nonumber\\
  &=&-\chi (N-N_{f}) \int\frac{d^{4}p}{(2\pi)^{4}}\frac{m_{\lambda}^{*}}{p^{2}
    +m_{\lambda}^{2}}+F^{*}
\end{eqnarray}
Here we set:
\begin{eqnarray}
  \chi&=&\frac{g^{2}N_{f}}{32\pi^{2}}\frac{1}{<\phi>}\nonumber\\
  m_{\lambda}&=&\chi F_{\phi}\nonumber
\end{eqnarray}

The extremum condition is $\frac{\partial V}{\partial F}=0$,
and this becomes a mass-gap equation:

\begin{eqnarray}
  \label{gape}
  1-\chi^{2}(N-N_{f})\int \frac{d^{4}p}{(2\pi)^{4}}\frac{1}{p^{2}
    +m_{\lambda}^{2}}=0
\end{eqnarray}

Taking $N_{f}$ and $N$ enough large,
 we can find that the condensation 
really occurs.

\section{Conclusion}
\hspace*{\parindent}
We have applied Nambu-Jona-Lasinio method to the supersymmetric SQCD
with the $SU(N)$ gauge group and  $N_{f}$ flavors.
It is disappointing that our argument cannot be applied
unless we take $N_{f}$ and $N$ enough large, but 
it is still important to show that the higher order interaction
term can induce a gaugino condensation in the perturbative
region.
Application of Nambu-Jona-Lasinio method to supergravity theory
is already studied in ref.\cite{ross} from another point of view.

\section*{Acknowledgment}
\hspace*{\parindent}
We thank K.Fujikawa and A.Yamada for many helpful discussions.

\end{document}